\documentclass[12pt]{article}
\usepackage{epsfig, epsf, graphicx, subfigure, epstopdf}
\usepackage{pstricks, pst-node, psfrag}
\usepackage{amssymb,amsmath,bm}
\usepackage{verbatim,enumerate}
\usepackage{rotating, lscape}
\usepackage{setspace}
\usepackage{multirow}
\usepackage{color}
\usepackage{animate}
\usepackage{multimedia}
\usepackage[normalem]{ulem}
\usepackage{lineno}

\usepackage{natbib}

\def\log{\hbox{log}}

\def\boxit#1{\vbox{\hrule\hbox{\vrule\kern6pt
          \vbox{\kern6pt#1\kern6pt}\kern6pt\vrule}\hrule}}

\def\bse{\begin{eqnarray*}}
\def\ese{\end{eqnarray*}}
\def\be{\begin{eqnarray}}
\def\ee{\end{eqnarray}}
\def\bq{\begin{equation}}
\def\eq{\end{equation}}
\def\bse{\begin{eqnarray*}}
\def\ese{\end{eqnarray*}}

\newcommand{\bB}{\mathbf{B}}

\newcommand{\bU}{\mathbf{U}}
\newcommand{\bI}{\mathbf{I}}

\newcommand{\bW}{\mathbf{W}}
\newcommand{\bX}{\mathbf{X}}
\newcommand{\bY}{\mathbf{Y}}
\newcommand{\bZ}{\mathbf{Z}}

\newcommand{\by}{\mathbf{y}}

\newcommand{\bu}{\mathbf{u}}
\newcommand{\bx}{\mathbf{x}}
\newcommand{\bz}{\mathbf{z}}
\newcommand{\bmu}{\boldsymbol{\mu}}
\newcommand{\blambda}{\boldsymbol{\lambda}}
\newcommand{\bomega}{\boldsymbol{\omega}}
\newcommand{\bOmega}{\boldsymbol{\Omega}}

\newcommand{\bxi}{\boldsymbol{\xi}}
\newcommand{\bdelta}{\boldsymbol{\delta}}

\newcommand{\bvepsilon}{\boldsymbol{\varepsilon}}
\newcommand{\balpha}{\boldsymbol{\alpha}}

\newcommand{\bseta}{\boldsymbol{\eta}}
\newcommand{\bzeta}{\boldsymbol{\zeta}}
\newcommand{\bDelta}{\boldsymbol{\Delta}}
\newcommand{\bSigma}{\boldsymbol{\Sigma}}

\newcommand{\bPsi}{\boldsymbol{\Psi}}
\newcommand{\bUpsilon}{\boldsymbol{\Upsilon}}

\newcommand{\btheta}{\boldsymbol{\theta}}
\newcommand{\0}{\mathbf{0}}
\newcommand{\1}{\mathbf{1}}
\newcommand{\bsy}{\boldsymbol}
\newcommand{\mbf}{\mathbf}

\newtheorem{myprop}{Proposition}

\newcommand{\iid}{\stackrel{\mathrm{iid}}{\sim}}

\begin{document}


\thispagestyle{empty} \baselineskip=28pt \vskip 5mm
\begin{center} {\huge{\bf A Multi-Resolution Spatial Model for Large Datasets Based on the Skew-$t$ Distribution}}
\end{center}

\baselineskip=12pt \vskip 10mm

\begin{center}\large
Felipe Tagle\footnote[1]{\baselineskip=10pt Department of Applied and Computational Mathematics and Statistics, University of Notre Dame, Notre Dame, IN 46556, United States.  \\ E-mail: ftagleso@nd.edu,  scastruc@nd.edu}, Stefano Castruccio$^{1}$ and Marc G.~Genton\footnote[2]{\baselineskip=10pt CEMSE Division, King Abdullah University of Science and Technology, Thuwal 23955-6900, Saudi Arabia. \\ E-mail: marc.genton@kaust.edu.sa \\
This publication is based upon work supported by the King Abdullah University of Science and Technology (KAUST) Office of Sponsored Research (OSR) under Award No: OSR-2015-CRG4-2640.}\\
\end{center}

\baselineskip=17pt \vskip 10mm \centerline{\today} \vskip 10mm

\begin{center}
{\large{\bf Abstract}}
\end{center}
\baselineskip=17pt

Large, non-Gaussian spatial datasets pose a considerable modeling challenge as the dependence structure implied by the model needs to be captured at different scales, while retaining feasible inference. Skew-normal and skew-$t$ distributions have only recently begun to appear in the spatial statistics literature, without much consideration, however, for the ability to capture dependence at multiple resolutions, and simultaneously achieve feasible inference for increasingly large data sets. This article presents the first multi-resolution spatial model inspired by the skew-$t$ distribution, where a large-scale effect follows a multivariate normal distribution and the fine-scale effects follow a multivariate skew-normal distributions. The resulting marginal distribution for each region is skew-$t$, thereby allowing for greater flexibility in capturing skewness and heavy tails characterizing many environmental datasets. Likelihood-based inference is performed using a Monte Carlo EM algorithm. The model is applied as a stochastic generator of daily wind speeds over Saudi Arabia.


\clearpage\pagebreak\newpage \pagenumbering{arabic}
\baselineskip=26.5pt

\section{Introduction}\label{sec:intro}


Along with advances in computational processing and storage capabilities, spatio-temporally referenced datasets in the geophysical and environmental sciences have been steadily increasing in size. As observations are simulated or recorded at finer temporal and spatial scales, potentially allowing for a host of new scientific questions to be answered, inferential aspects of even the simplest of geostatistical models becomes problematic. Indeed, datasets have reached sizes that are orders of magnitude larger than those that classic statistical theory was designed to deal with \citep{efron2016computer}. Gaussian process models underpin much of the spatial statistics literature. Despite the analytical tractability of the Gaussian distribution, the application of these models to large datasets becomes computationally challenging since the evaluation of the likelihood at $n$ spatial locations requires generally $\mathcal{O}(n^3)$ operations and $\mathcal{O}(n^2)$ in memory \citep{sun2012geostatistics}. Thus, inference quickly becomes difficult for datasets indexed at $n=100,000$ locations, which may be considered of medium size in global climate model output. 

The need for new methodologies to address these inferential challenges has led to the development of several modeling approaches, such as sparse approximations of the covariance function \citep{furrer2006covariance,kaufman2008covariance}, imposing separability \citep{genton2007separable}, the use composite likelihoods \citep{stein2004approximating} and localizing the dependence structure through Gaussian Markov random fields \citep{lindgren2011explicit}. Another line of research has placed less emphasis on the second moment structure and has focused on explicitly modeling the dynamics of the spatio-temporal process \citep{wikle2010general,wikle2015modern}, which in the context of large datasets typically employs reduced rank approximations of the underlying process \citep{cressie2008fixed,katzfuss2012bayesian,sengupta2013hierarchical}. The approximations are motivated by the Karhunen-Lo{\`e}ve expansion of a stochastic process \citep{adler2010geometry}, which states that under mild regularity conditions it may be decomposed into a countable orthonormal series, which can then be truncated to yield a finite approximation. \citet{sang2012full} combined sparse matrix methods and a low-rank approximation to the covariance function to capture both large and small spatial scale variation. The Gaussian assumption, however, is seldom justified in applications, as observations typically exhibit departures from Gaussianity in the form of heavy-tails and/or skewness. The standard approach to model such data involves applying a transformation to the original data with the aim of inducing an empirical distribution that more closely resembles the Gaussian \citep[e.g.,][]{xu2017tukey}. Another, arguably more natural, approach is to discard the assumption and use a class of distributions that offers the flexibility of explicit modeling the higher moments of the spatial process \citep{roislien2006t,bolin2014spatial}.

In the seminal work of \citet{azzalini1985class}, the class of skew-normal distributions was introduced, which extended the normal distribution to incorporate varying degrees of skewness. At its foundation lies the program of perturbing or modulating a symmetric probability density by a factor corresponding to a continuous distribution function on the real line. This general construction leads naturally in the multivariate setting to the skew-elliptical distributions, a particular example of which is the skew-$t$ distribution, an extension of the Student-$t$ distribution designed to capture both skewness and excess kurtosis \citep{sahu2003new, azzalini2003distributions}. Recently, the multivariate skew-$t$ (ST) distribution has been employed extensively in applied studies \citep{thompson2004coastal, tagle2017assessing} and more recently in mixture modelling \citep{lin2010robust,lee2011fitting}; see also \citep{genton2004skew} for other applications.

The additional flexibility offered by the ST distribution comes at the expense of several desirable properties that characterize the normal distribution, namely, closure under convolutions and conditioning. This poses difficulties in applications where, for instance, processes are represented as sums of subprocesses operating at different scales. In geophysics, \citet{wikle2001spatiotemporal} proposed modeling tropical surface wind fields as the sum of two processes, one representing large-scale atmospheric phenomena and another capturing fine-scale motion. Similarly, in medical imagery, \citet{castruccio2017multi} proposed a multi-resolution spatio-temporal model for brain activation using fMRI data, allowing for local non-stationary spatial dependence within anatomically defined regions of interest (ROIs) as well as regional dependence across ROIs. These examples underscore the need to extend the ST, allowing for scalable and flexible models that are able to capture features of multi-resolution processes.

In this work, we propose a methodology that represents the first attempt to extend the multivariate ST to a multi-resolution framework. It exploits the construction of this distribution as a multivariate skew-normal distribution (SN) rescaled by the square root of a $\chi^2$ distribution, and the closure of the former under convolutions with a normal distribution \citep{azzalini2014skew}. In particular, we consider 2-level hierarchy, where a large-scale effect interacts linearly with fine-scale regional processes; the former is assumed to follow a multivariate normal distribution, with each component operating in a different designated region; while the regional processes each follow independent multivariate SN distributions. The construction results in within- and across-region dependent marginal ST distributions, and at the same time, the regionalization of the spatial domain with independent components offers a flexibility that is aptly suited for large datasets. A convenient stochastic representation of the SN distribution \citep{azzalini1996multivariate} is used to develop an Expectation-Maximization (EM) algorithm \citep{dempster1977maximum} for likelihood-based inference. Closed form expressions are derived for several of the expressions of the E-step and M-step, while for those that were analytically intractable a Monte Carlo approach was used.

We consider a discrete spatial domain, thus we are not concerned with the technicalities of continuous stochastic processes. Consequently, our focus is not on the traditional application of spatial interpolation or kriging, where a partial observation of such a spatial process is observed and a prediction of its value at an unobserved location is desired. Instead, we focus on parameter estimation and envision the fitted model being used to generate synthetic replicates of the training dataset, an example of which is provided in the Application section where it is used as a stochastic generator \citep{jeo17a} of daily wind fields.

The remainder of the paper is organized as follows: Section \ref{sec:prelim} presents a few results related to the skew-normal and skew-$t$ distributions that are germane to the study; Section \ref{sec:model} describes the construction of the multi-resolution model, as well as the detail of the EM algorithm used for inference; Section \ref{sec:SimStudy} presents a simulation study that compares the performance against a Gaussian model; Section \ref{sec:Application} provides an  application that builds on the work of \citet{tagle2017assessing} on wind fields over Saudi Arabia and Section \ref{sec:Conclusions} concludes.

\section{Preliminaries}\label{sec:prelim}

In this section we provide an introduction to the multivariate SN and ST distributions. A random $d$-dimensional vector $\bY$ follows a standard multivariate SN distribution \citep{azz99}, denoted as $SN_d(\0,\bar{\bOmega},\bsy{\alpha})$, if it has a probability density of the form
\begin{equation}\label{eq:SN}
2 \phi_d(\by; \bar{\bOmega}) \Phi(\balpha^\top \by) ,\quad \by \in \mathbb{R}^d,
\end{equation}
where $\bar{\bOmega}$ is a $d \times d$ correlation matrix, $\phi_d(\by;\bSigma)$ denotes the $d$-dimensional probability density function of a $N_d(\0, \bSigma)$ variate, $\Phi(\cdot)$ denotes the cumulative distribution function of a standard normal variate, and $\balpha$ is a $d$-dimensional skewness parameter. If $\balpha=\0$, then \eqref{eq:SN} reduces to a standard $d$-dimensional normal random variate, whereas $\balpha \neq \0$ generates an asymmetric family of distributions. 

Similarly, a random $d$-dimensional vector $\bY$ follows a normalized multivariate ST distribution \citep{azzalini2003distributions}, if it has a probability density of the form 
\begin{equation}\label{eq:ST}
2t_d(\by; \bar{\bOmega},\nu) T\left(\balpha^\top \by \sqrt{\frac{ \nu + d}{\nu + Q(\bz)}}; \nu + d \right),
\end{equation}
where $\bar{\bOmega}$ is defined analogously, $Q(\by) = \by^\top \bar{\bOmega} \by$, $t_d(\cdot; \bar{\bOmega}, \nu)$ is the probability density function of the $d$-dimensional Student-$t$ distribution, and $T(\cdot; \nu)$ denotes the cumulative distribution function of a univariate Student-$t$ distribution with $\nu$ degrees of freedom. It is the natural extension of the multivariate Student-$t$ distribution, given by the ratio $\bY = \bX/\sqrt{Z}$, where $Z \sim \chi^2_\nu/\nu$, with $\chi^2_\nu$ denoting a chi-squared distribution with $\nu$ degrees of freedom, and $\bX$ an independent $N_d(\0,\bSigma)$ variate. In contrast, the construction of the multivariate skew-$t$ distribution assumes that $\bX$ is a multivariate SN distribution. As $\nu \to \infty$, the density function \eqref{eq:ST} converges to \eqref{eq:SN}. The first moment exists if $\nu > 1$ otherwise, if  $\nu = 1$, \eqref{eq:ST} reduces to the skew-Cauchy distribution \citep{arn00}. Given a location parameter $\bxi$ and scale matrix $\bomega$, the transformation $\bxi + \bomega Y$ gives rise to the $d$-dimensional multivariate ST distribution denoted by $ST_d(\bxi,\bOmega, \balpha,\nu)$, with $\bOmega = \bomega \bar{\bOmega} \bomega$. 

One of the difficulties of using the ST distribution and its multivariate counterpart in applications is its lack of closure under addition. We can prove, however, that the scaled sum of a normal and a SN does belong to the ST family; the proof is found in the appendix.

\begin{myprop}
Let $X_0 \sim N(0,1)$, $\bX \sim SN_d(\0,\bOmega,\balpha)$, $Z \sim Gamma(\nu/2,\nu/2)$, independent, $\bzeta = (\zeta_1,\ldots,\zeta_d)^{\top} \in (-1,1)^d$, and $\bDelta_{\zeta} = \left( \bI_d - \text{diag}(\bzeta)^2 \right)^{1/2}$ then
\begin{equation}\label{skew_r}
\bY = \frac{\bzeta X_0 + \bDelta_{\zeta} \bX}{\sqrt{Z}} 
\end{equation}
has distribution $ST_d(\0,\bOmega^*, \balpha^*,\nu)$, with
\begin{equation*}
\bOmega^* = \bDelta (\bOmega + \blambda \blambda^\top) \bDelta, \quad  \balpha^* = \left( 1 + \frac{\balpha^\top \blambda \blambda^\top \balpha}{1 + \blambda^\top \bOmega^{-1} \blambda} \right)^{-1/2} {\bOmega^*}^{-1}\bDelta \bOmega \balpha,
\end{equation*}
where $\blambda = (\lambda(\zeta_1),\ldots, \lambda(\zeta_d))^\top$, and $\lambda(\zeta) = \zeta/(1-\zeta^2)^{1/2}$.
\end{myprop}

\section{Multi-resolution Skew-$t$ Model}\label{sec:model}

\subsection{Model definition}

We consider a multi-resolution model in which a large-scale effect $X_0$ interacts with a $d$-vector of fine-scale effects $\bX$. We consider the stochastic representation of the multivariate SN distribution, $\bX = \bdelta |U_0| + \bDelta_{\delta} \bU$,
where $U_0 \sim N(0,1)$ and $\bU \sim N_d(\0,\bPsi)$ are independent, with $\bdelta = (\delta_1,\ldots,\delta_d)^{\top}$, and $\bDelta_{\delta}=\left( \bI_d - \text{diag}(\bdelta)^2 \right)^{1/2}$  \citep{azzalini1996multivariate}. By plugging this representation into \eqref{skew_r}, we obtain
\[ \bY = \frac{\bzeta X_0 + \bDelta_{\zeta} \left( \bdelta |U_0| + \bDelta_{\delta} \bU \right)}{\sqrt{Z}}. \]
which is $ST_d$ distributed. 

Another approach to the construction of multi-resolution models for large datasets is based on the use of multiple sets of fixed basis functions, each set intended to represent a different scale of spatial variation \citep{cressie2008fixed,katzfuss2012bayesian}. Random variation is achieved by coupling such functions with a lower-dimensional Gaussian process. The choice of basis functions offers the flexibility to capture any non-stationarities across the spatial domain, and it remains an area of research \citep[e.g.][]{bradley2011selection}. Our construction admits only two resolutions, where at a first stage the spatial domain is partitioned into a collection of regions where the assumption of stationarity is plausibly valid, thereby allowing the well-developed theory of stationary covariance functions to be exploited; and at a second stage, another stationary process is added that interacts linearly with the first. In particular, we assume $\bY=(\bY_1^{\top}, \ldots, \bY_R^{\top})^{\top}$, i.e., the vector of observations is split into $R$ regions of size $d_r$, with $\sum_{r=1}^R d_r=d$, and the former expression can be applied to every region separately, so that
\begin{equation}\label{eq:Yr_rep}
\bY_r = \frac{\bzeta_r X_{0,r} + \bDelta_{\zeta_r} \bdelta_r |U_{0,r}| + \bDelta_{\zeta_r,\delta_r} \bU_r}{\sqrt{Z_r}}, \quad r=1,\ldots,R,
\end{equation}
where $\bDelta_{\zeta_r,\delta_r} = \bDelta_{\zeta_r} \odot \bDelta_{\delta_r}$, with $\odot$ denoting the Hadamard product. Here, $U_{0,r}\iid N(0,1)$, $Z_r$ independent with distribution Gamma$(\nu_r/2,\nu_r/2)$, $\bU_r$ independent with distribution $N_d(\0,\bPsi(\btheta_{\bU_r}))$. Spatial dependence across regions is introduced through the large-scale random vector $\bX_0 = (X_{0,1},\ldots,X_{0,R})^\top\sim N_R(\0,\bSigma(\btheta_{\bX_0}))$. Hence, $(X_{0,r}, U_{0,r}, Z_r)^\top$, $r=1, \ldots, R,$ are vectors of latent processes in \eqref{eq:Yr_rep}, and the parameters of the model are $\btheta=(\{\bdelta_r^{\top},\bzeta_r^{\top},\nu_r^{\top},\btheta_{\bU_r}^{\top}, r=1, \ldots, R\},\btheta_{\bX_0}^{\top})^{\top}$. For notational simplicity, we drop the parametric dependence of the covariance matrices and denote $\bPsi_r=\bPsi(\btheta_{\bU_r})$ and $\bSigma=\bSigma(\btheta_{\bX_0})$. 

The representation in \eqref{eq:Yr_rep} makes an EM algorithm \citep{dempster1977maximum} a natural choice for inference. It aims to maximize the model likelihood in the presence of latent processes, by alternating between an expectation  and a maximization step, see \citet{mclachlan2007algorithm} for details. During the first step, $(X_{0,r}, U_{0,r}, Z_r)^\top, r=1, \ldots, R$ and functions thereof are replaced by their conditional expectations given the data and parameter estimates $\hat{\btheta}$, and in the second, $\hat{\btheta}$ is updated in the traditional maximum-likelihood sense based on the maximization of the associated log-likelihood. Since a parameter set is required to conduct the first E-step, an initial estimate denoted by $\hat{\btheta}_0$ is provided, which is then subsequently updated by alternating between both steps until convergence is achieved.

\subsection{EM algorithm}

Let $\eta_{0,r} = X_{0,r}/\sqrt{Z_r}$ and $\eta_{1,r} = |U_{0,r}|/\sqrt{Z_r}$, $\bseta_0 = (\eta_{0,1},\ldots,\eta_{0,R})^\top$, with $\bseta_1$ and $\bZ$ defined analogously, so that the latent processes are $\{\bseta_{0}, \bseta_{1}, \bZ\}$. Then $\bY_r$, $r=1,\ldots,R$ can be represented hierarchically as:
\[
\begin{array}{lll}
\bY_r | \eta_{0,r},\eta_{1,r},Z_r & \overset{iid}{\sim} & N_{d_r}\left(\bzeta_r \eta_{0,r}  + \bDelta_{\zeta_r} \bdelta_r \eta_{1,r}, \frac{1}{Z_r} \bUpsilon_r \right), \\
\bseta_{0} |  \bseta_{1}, \bZ   & \sim & N_R(\0,\bDelta_{z}^\top \bDelta_{z} ), \\
\eta_{1,r} |  Z_r   & \overset{iid}{\sim} & HN(0,1/Z_r), \\
Z_r & \overset{iid}{\sim} & \text{Gamma}(\nu_r/2,\nu_r/2), 
\end{array}
\]
\begin{sloppypar}
\noindent where $HN$ refers to the half-normal distribution, $\bUpsilon_r = \bDelta_{\zeta_r,\delta_r}^\top \bPsi_r \bDelta_{\zeta_r,\delta_r}$, and $\bDelta_{z} =\text{diag}\left(1/\sqrt{Z_{1}},\ldots, 1/\sqrt{Z_{R}}\right)$. Similar hierarchical representations can be found in other EM algorithm implementations based on the SN distribution \citep[e.g.,][]{arellano2005skew,lin2008estimation, lachos2010likelihood}. 
\end{sloppypar}

Assuming a collection of observations in time $\by_t = (\by_{1,t}^\top,\ldots\by_{R,T}^\top)^\top$, $t=1,\ldots,T$, we can express the joint distribution for each $t$ as
\begin{equation}\label{eq:joint}
\begin{split}
p(\by_t, \bseta_{0,t},\bseta_{1,t},\bZ_t \mid \bsy{\theta})  = & \prod_{r=1}^R \frac{(\nu_r/2)^{\nu_r/2}}{\Gamma(\nu_r/2)}\frac{1}{(2\pi)^{d_r/2}} \sqrt{\frac{2}{\pi}} \frac{1}{|\bUpsilon_r|^{1/2}} \;  Z_{r,t}^{\frac{\nu_r+d_r}{2}}
\exp\left\{ -\frac{Z_{r,t}}{2} \left[ \bx_{r,t}^\top \bUpsilon_r^{-1}  \bx_{r,t} +\eta_{1,r,t}^2 + \nu_r  \right] \right\} \\
& \times \frac{1}{(2\pi)^{R/2}} \frac{1}{|\bSigma|^{1/2}} \exp\left\{-\frac{1}{2}\bseta_{0,t}^\top \bDelta_{z,t}^{-1} \bSigma^{-1} \bDelta_{z,t}^{-1} \bseta_{0,t} \right\},
\end{split}
\end{equation}
where now $\bseta_{0,t} = (\eta_{0,1,t},\ldots,\eta_{0,R,t})^\top$, $\bseta_{1,t}$ and $\bZ_t$ are defined analogously,  $\bx_{r,t} = \by_{r,t} - \bzeta_r \eta_{0,r,t} -  \bDelta_{\zeta_r} \bdelta_r \eta_{1,r,t}$, and $\bDelta_{z,t} =\text{diag}(1/\sqrt{Z_{1,t}},\ldots, 1/\sqrt{Z_{R,t}})$. If the total vector of observations is $\by=(\by_1^{\top},\ldots,\by_T^{\top})^{\top}$ (with similar notation for each of the latent processes $\{\bseta_0, \bseta_1,\bZ\}$, the corresponding log-likelihood for all points, excluding additive constants, is
\begin{equation}\label{log_comp}
\boldsymbol{\ell}(\btheta | \by, \bseta_0, \bseta_1,\bZ)  = \sum_{r=1}^R \boldsymbol{\ell}_r(\btheta | \by, \bseta_0, \bseta_1,\bZ)   - \frac{T}{2}\log|\bSigma|    -  \frac{1}{2} \sum_{t=1}^T \bseta_{0,t}^\top \bDelta_{z,t}^{-1} \bSigma^{-1} \bDelta_{z,t}^{-1} \bseta_{0,t},
\end{equation}
where
\begin{equation*}
\begin{split}
\boldsymbol{\ell}_r(\btheta | \by, \bseta_0, \bseta_1, \bZ)  = &  \; T\frac{\nu_r}{2} \log \left(\frac{\nu_r}{2}  \right) - T\log\, \Gamma \left(\frac{\nu_r}{2} \right)  - \frac{T}{2} \log |\bUpsilon_r|   \\ 
& +\frac{\nu_r+d_r}{2} \sum_{t=1}^T \log(Z_{r,t}) - \sum_{t=1}^T \frac{Z_{r,t}}{2}  \left[ \bx_{r,t}^\top \bUpsilon_r^{-1}  \bx_{r,t} +\eta_{1,r,t}^2 + \nu_r \right]   \\
= &  \; T\frac{\nu_r}{2} \log \left(\frac{\nu_r}{2}  \right) - T\log\, \Gamma \left(\frac{\nu_r}{2} \right) -  \frac{T}{2}\sum_{s=1}^{d_r}\log (1-\zeta_{r,s}^2) - \frac{T}{2}\sum_{s=1}^{d_r}\log (1-\delta_{r,s}^2) - \frac{T}{2} \log |\bPsi_r|   \\ 
& +\frac{\nu_r+d_r}{2} \sum_{t=1}^T \log(Z_{r,t})  - \frac{1}{2}\sum_{t=1}^T Z_{r,t} \left[ \bx_{r,t}^\top \bDelta_{\zeta_r,\delta_r}^{-1} \bPsi_r^{-1} \bDelta_{\zeta_r,\delta_r}^{-1}\bx_{r,t} + \eta_{1,r,t}^2 + \nu_r \right]. 
\end{split}
\end{equation*}
The log-likelihood involves the inversion of the matrices $\bPsi_r$, and $\bSigma$, as well as the computation of their respective determinants, which are typically problematic when the size of dataset becomes large. However, our approach of regionalizing the spatial domain ensures that the size of each $\bPsi_r$ remains within limits of computational feasibility, while the dimension of $\bSigma$ is limited to the number of regions $R$, for which $R \ll  d$ is assumed to hold. Furthermore, the independence assumption across the regions allows for further computational efficiency through parallelization of the respective operations.

\subsection{E-step}\label{subsec:ModelE-step}

Let us assume that we have $\btheta = \hat{\btheta}^{[k]}$ at the $k$-th iteration. From \eqref{log_comp}, we can compute the  $Q(\btheta | \hat{\btheta}^{[k]}) = \mathbb{E} \left[\boldsymbol{\ell}(\btheta | \by, \bseta_0, \bseta_1,\bZ) | \by, \hat{\btheta}^{[k]}\right]$,
which can be simplified into
\begin{equation*}
Q(\btheta | \hat{\btheta}^{[k]})  = \sum_{r=1}^R \langle \boldsymbol{\ell}_r(\btheta | \by, \bseta_0, \bseta_1,\bZ) | \by, \hat{\btheta}^{[k]} \rangle - \frac{T}{2}\log|\bSigma|    -  \frac{1}{2} \text{tr} \left\{ \bSigma^{-1} \sum_{t=1}^T \langle (\bseta_{0,t} \odot \sqrt{\bZ_{t}}) (\bseta_{0,t} \odot \sqrt{\bZ_{t}} )^\top \rangle \right\}
\end{equation*}
where $\langle \cdot \rangle := \mathbb{E}(\cdot | \by, \hat{\btheta}^{(k)})$, and the conditional expectation of the Hadamard product is understood to be component-wise. Furthermore, if for simplicity we denote by $ \boldsymbol{\ell}_r=\boldsymbol{\ell}_r(\btheta | \by, \bseta_0, \bseta_1,\bZ)$:
\begin{equation}
\begin{split}
\mathbb{E} \left[ \boldsymbol{\ell}_r | \by, \hat{\btheta}^{[k]} \right] =  & \; T\frac{\nu_r}{2} \log \left(\frac{\nu_r}{2}  \right) - T\log\, \Gamma \left(\frac{\nu_r}{2} \right)  - \frac{T}{2}\sum_{s=1}^{d_r}\log (1-\zeta_{r,d}^2) - \frac{T}{2}\sum_{s=1}^{d_r}\log (1-\delta_{r,d}^2)  -  \frac{T}{2} \log |\bPsi_r|  \\ 
& +\frac{\nu_r+d_r}{2} \sum_{t=1}^T \langle\log(Z_{r,t})\rangle - \frac{1}{2} \sum_{t=1}^T \langle Z_{r,t} \eta_{r,t}^2 \rangle  - \frac{\nu_r}{2}\sum_{t=1}^T \langle Z_{r,t} \rangle  \\
&  -\frac{1}{2} \text{tr} \left\{ \bPsi_r^{-1} \bDelta_{\zeta_r,\delta_r}^{-1} \bB_r \bDelta_{\zeta_r,\delta_r}^{-1}\right\},
\end{split}\label{eq:ExpReg}
\end{equation}
and
\begin{equation*}
\begin{split}
\bB_r = & \sum_{t=1}^T \langle Z_{r,t} \rangle  \by_{r,t} \by_{r,t}^\top  - \sum_{t=1}^T \langle Z_{r,t}  \eta_{0,r,t}\rangle \bzeta_r\by_{r,t}^\top  -  \sum_{t=1}^T \langle Z_{r,t} \eta_{0,r,t}\rangle \by_{r,t}\bzeta_r^\top  \\ 
&  -\sum_{t=1}^T \langle Z_{r,t} \eta_{1,r,t}\rangle  \bDelta_{\zeta_r}\bdelta_r \by_{r,t}^\top - \sum_{t=1}^T \langle Z_{r,t} \eta_{1,r,t}\rangle \by_{r,t} \bdelta_r ^\top \bDelta_{\zeta_r}  + \sum_{t=1}^T \langle Z_{r,t} \eta_{0,r,t}^2\rangle \bzeta_r\bzeta_r^\top    \\
 & +\sum_{t=1}^T \langle Z_{r,t} \eta_{0,r,t}\eta_{1,r,t} \rangle \bDelta_{\zeta_r} \bdelta_r\bzeta_r^\top + \sum_{t=1}^T \langle Z_{r,t} \eta_{0,r,t}\eta_{1,r,t} \rangle \bzeta_r\bdelta_r^\top \bDelta_{\zeta_r} + \sum_{t=1}^T \langle Z_{r,t} \eta_{1,r,t}^2\rangle  \bDelta_{\zeta_r}\bdelta_r\bdelta_r^\top \bDelta_{\zeta_r}. 
\end{split}
\end{equation*}
The intractability of the above conditional expectations suggests the use of a Monte Carlo EM approach \citep{wei1990monte}, in particular, for each $t$, and iteration $k$, we generate vectors $(\bseta_{0,t}^{(m)\top},\bseta_{1,t}^{(m)\top},\bZ_{t}^{(m)\top})^\top$, $m=1,\ldots,M$ from $p(\bseta_{0,t},\bseta_{1,t},\bZ_t | \by_t, \hat{\btheta}^{[k]})$ (which is derived from \eqref{eq:joint}) using a Gibbs sampler.  Thus for the functions above, denoted generically by $g$, we have  $\langle g(Z_{r,t},\eta_{0,r,t},\eta_{1,r,t}) \rangle = \frac{1}{M} \sum_{m=1}^M g(Z_{r,t}^{(m)},\eta_{0,r,t}^{(m)},\eta_{1,r,t}^{(m)})$. Because the random vectors are independent across time, the sampling from $p(\bseta_{0,t},\bseta_{1,t},\bZ_t | \by_t, \hat{\btheta}^{[k]})$ admits a straightforward parallelization that greatly reduces the computational burden of employing a Monte Carlo-based method. Otherwise, the conditional distributions could have been approximated using the Laplace method \citep{sengupta2013hierarchical}. Throughout this section, we denote with the hat notation the parameters estimated from the M-step at the previous iteration, and we drop the index $k$ for simplicity. The Gibbs sampler proceeds as follows: first we initialize $(Z_{r,t}^{(1)},\eta_{0,r,t}^{(1)},\eta_{1,r,t}^{(1)})^\top$, then for each $t$, $r$, and $m=2, \ldots, M$
\vspace{.1in}

\noindent 1) The full conditional distribution of $Z_{r,t}^{(m)}$ is not available in closed-form, thus we use a independent Metropolis-Hastings step within the Gibbs sampler. For each region $r$, we draw a sample from a candidate distribution $q(Z_r) \iid \text{Gamma}(\alpha_r,\beta_r)$ across $t$ and $m$, where 
\[ \alpha_r = \frac{\hat{\nu}_r + d_{r}}{2} +1,\qquad \beta_r = \frac{1}{2} \left(\bx_{r,t}^{(m)\top} \hat{\bUpsilon}_r^{-1}  \bx_{r,t}^{(m)}  + \hat{\nu}_r \right), \]
and $\bx_{r,t}^{(m)} = \by_{r,t} - \hat{\bzeta}_r \eta^{(m-1)}_{0,r,t} -  \bDelta_{\hat{\zeta}_r} \hat{\bdelta}_r \eta^{(m-1)}_{1,r,t}$.

The target distribution here follows from Eq.~\eqref{eq:joint}, 
\[ \pi(Z_{r,t}) \propto  Z_{r,t}^{\frac{\nu_r+d_r}{2}}
\exp\left\{ -\frac{Z_{r,t}}{2} \left[ \bx_{r,t}^\top \hat{\bUpsilon}_r^{-1}  \bx_{r,t} +\eta_{1,r,t}^2 + \hat{\nu}_r  \right]  - \frac{1}{2} \left[ Z_{r,t} \eta_{0,r,t}^2 \hat{\lambda}_{r,r} + 2 \sum_{j\neq r} \sqrt{Z_{r,t}Z_{j,t}} \eta_{0,r,t}\eta_{0,j,t} \hat{\lambda}_{r,j}  \right] \right\}, \]
where $\hat{\lambda}_{i,j} = (\hat{\bSigma}^{-1})_{i,j}$. 

This sampler considers an acceptance probability given by $\alpha_{prob}(Z_{r,t}^{(m-1)},Y) = \min \left\{1,\frac{w\left(Y\right)}{w\left(Z_{r,t}^{(m-1)}\right)}\right\}$, where $w(\cdot) = \pi(\cdot)/q(\cdot)$, i.e., the ratio of the target and candidate distribution, in our case
\[ w(Z_{r,t}) = \exp\left\{-\frac{Z_{r,t}}{2}\eta_{1,r,t}^2 - \frac{1}{2} \left[ Z_{r,t} \eta_{0,r,t}^2 \hat{\lambda}_{r,r} + 2 \sum_{j\neq r} \sqrt{Z_{r,t}Z_{j,t}} \eta_{0,r,t}\eta_{0,j,t} \hat{\lambda}_{r,j}  \right] \right\}. \] 

Thus we generate $Y\sim q$ and given $U \sim U(0,1)$, if $U < \alpha_{prob} (Z_{r,t}^{(m-1)},Y)$, $Z_{r,t}^{(m)} = Y$, otherwise $Z_{r,t}^{(m)} = Z_{r,t}^{(m-1)}$.

\vspace{.1in}

\noindent 2) Generate $\eta_{0,r,t}^{(m)}$ from $p\left(\eta_{0,r,t}^{(m)}| \bseta_{0,-r,t}^{(m)}, \bseta_{1,t}^{(m-1)}, \bz_t^{(m)}, \by_{r,t}\right) \sim  N\left(\mu_{0,r,t}^{(m)},\sigma_{0,r,t}^{(m)\,2}\right)$, where  
\[ 
\begin{array}{lll}
\bseta_{0,-r,t}^{(m)} & = & \left(\eta_{0,1,t}^{(m)},\ldots,\eta_{0,r-1,t}^{(m)},\eta_{0,r+1,t}^{(m-1)},\ldots,\eta_{0,R,t}^{(m-1)}\right), \\[7pt]
\mu_{0,r,t}^{(m)} & = &  \frac{ \hat{\bzeta}_r^\top \hat{\bDelta}_{\zeta_r,\delta_r}^{-1} \hat{\bPsi}_r^{-1} \hat{\bDelta}_{\zeta_r,\delta_r}^{-1}  (\by_{r,t} - \eta_{1,r,t}^{(m)} \hat{\bDelta}_{\zeta_r}\hat{\bdelta}_r) - 1/\sqrt{Z_{r,t}^{(m)}} \sum_{j \neq r} \eta_{0,j,t}^{(k)}\sqrt{Z_{j,t}^{(m)}}\hat{\lambda}_{r,j}}{\hat{\bzeta}_r^\top \hat{\bDelta}_{\zeta_r,\delta_r}^{-1} \hat{\bPsi}_r^{-1} \hat{\bDelta}_{\zeta_r,\delta_r}^{-1}\hat{\bzeta}_r + \hat{\lambda}_{r,r}^2}, \\[7pt]
\sigma_{0,r,t}^{(m)\,2} & = & \frac{1}{Z_{r,t}^{(m)}\left(\hat{\bzeta}_r^\top \hat{\bDelta}_{\zeta_r,\delta_r}^{-1} \hat{\bPsi}_r^{-1} \hat{\bDelta}_{\zeta_r,\delta_r}^{-1}\hat{\bzeta}_r + \hat{\lambda}_{r,r}^2 \right)},
\end{array}
\]
with $\eta_{0,j,t}^{(k)} = \eta_{0,j,t}^{(m)}$ if $j<r$ and $= \eta_{0,j,t}^{(m-1)}$ otherwise.  

\vspace{.1in}
\noindent 3) Generate $\eta_{1,r,t}^{(m)}$ from $p\left(\eta_{1,r,t}^{(m)} | \bseta_{1,-r,t}^{(m)}, \bseta_{0,t}^{(m)}, \bz_t^{(m)}, \by_t \right) \sim  HN\left(\mu_{1,r,t}^{(m)},\sigma_{1,r,t}^{(m) 2}\right)$, with $ \bseta_{1,-r,t}^{(m)}$ defined analogously to $\bseta_{0,r-1,t}^{(m)}$ above,
\[ 
\mu_{1,r,t}^{(m)} =  \frac{\hat{\bdelta}_r^\top \hat{\bDelta}_{\zeta_r}\hat{\bDelta}_{\zeta_r,\delta_r}^{-1} \hat{\bPsi}_r^{-1} \hat{\bDelta}_{\zeta_r,\delta_r}^{-1}(\by_{r,t} - \eta_{0,r,t}^{(m)} \hat{\bzeta}_r )}{1+ \hat{\bdelta}_r^\top \hat{\bDelta}_{\zeta_r}\hat{\bDelta}_{\zeta_r,\delta_r}^{-1} \hat{\bPsi}_r^{-1} \hat{\bDelta}_{\zeta_r,\delta_r}^{-1} \hat{\bDelta}_{\zeta_r}\hat{\bdelta}_r }, \quad \sigma^{(m) 2}_{1,r,t} = \frac{1}{Z_{r,t}^{(m)}(1+\hat{\bdelta}_r^\top \hat{\bDelta}_{\zeta_r}\hat{\bDelta}_{\zeta_r,\delta_r}^{-1} \hat{\bPsi}_r^{-1} \hat{\bDelta}_{\zeta_r,\delta_r}^{-1} \hat{\bDelta}_{\zeta_r}\hat{\bdelta}_r)}.  \]


\vspace{.1in}
Because of the Monte Carlo errors introduced at this step, an increase in likelihood is not guaranteed at each iteration. We thus follow  the suggestions in \citet{mcculloch1994maximum} to increase the Monte Carlo sample size $M$ accordingly to quantify these errors. In practice, we choose $M$ as small as possible at the beginning of the algorithm and systematically increase $M$ with the number of iterations.  To assess the convergence of the algorithm, we consider Aitken acceleration-based stopping criterion \citep{ait26}.

\subsection{M-step}\label{subsec:ModelM-step}

\noindent 1) \textbf{Update $\nu_r^{(k)}$:} Obtained as solution to the following equation
\[ \log \left( \frac{\nu_r}{2} \right) + 1 - \text{DG}\left( \frac{\nu_r}{2} \right) + \frac{1}{T} \sum_{t=1}^T \left( \langle \log Z_{r,t} \rangle - \langle Z_{r,t}  \rangle \right) = 0, \]
with $\text{DG}$ the digamma function.

\noindent 2) \textbf{Update $\bPsi_r^{(k)}$:} The update is obtained numerically,
\[ \hat{\bPsi}_r^{(k+1)}  =  \hat{\bPsi}_r^{(k+1)} (\btheta_{\bU_r}^*), \quad \btheta_{\bU_r}^*= \text{argmax}_{\btheta_{\bU_r}} \left\{ - \frac{T}{2} \log |\bPsi_r(\btheta_{\bU_r})| - \frac{1}{2} \text{tr} \left\{ \bPsi_r(\btheta_{\bU_r})^{-1} \hat{\bDelta}_{\zeta_r,\delta_r}^{-1}  \hat{\bB}_r \hat{\bDelta}_{\zeta_r,\delta_r}^{-1}\right\} \right\}. \] 
If a non-parametric estimate is desired, it can be obtained from the first order condition applied to Eq. \eqref{eq:ExpReg}, namely,
\begin{equation*}
\hat{\bPsi}_r^{(k+1)} = \frac{1}{T} \hat{\bDelta}_{\zeta_r,\delta_r}^{-1}  \hat{\bB}_r \hat{\bDelta}_{\zeta_r,\delta_r}^{-1}.
\end{equation*}

\noindent 3) \textbf{Update $\bzeta_r^{(k)}$:} The update is obtained by numerically, 
\begin{equation*}
 \hat{\bzeta}_r^{(k+1)} = \text{argmax}_{\bzeta_r} \left\{ - \frac{T}{2}\sum_{s=1}^{d_r}\log (1-\zeta_{r,s}^2)  - \frac{1}{2} \text{tr} \left\{ \hat{\bPsi}_r^{-1} \bDelta_{\zeta_r,\delta_r}^{-1} \bB_r \bDelta_{\zeta_r,\delta_r}^{-1}\right\} \right\}. 
\end{equation*}
 \vspace{.1in}
 
\noindent 4) \textbf{Update $\bdelta_r^{(k)}$:}  The update is analogous to that of $\bzeta_r^{(k)}$,

\begin{equation*}
\hat{\bdelta}_r^{(k+1)} = \text{argmax}_{\bdelta_r} \left\{ - \frac{T}{2}\sum_{s=1}^{d_r}\log (1-\delta_{r,s}^2)  - \frac{1}{2} \text{tr} \left\{ \hat{\bPsi}_r^{-1} \bDelta_{\zeta_r,\delta_r}^{-1} \bB_r \bDelta_{\zeta_r,\delta_r}^{-1}\right\} \right\}. 
\end{equation*}

\noindent 5) \textbf{Update $\bSigma^{(k)}$:} Analogous to the case of $\bsy{\Psi}_r^{(k)}$, if the number of regions under considerations makes it appropriate to consider parameterizing $\bSigma$, then the update would be of the form, 
\begin{align*}
\hat{\bSigma}^{(k+1)}  &=  \hat{\bSigma}^{(k+1)} (\btheta_{\bSigma}^*), \\
 \btheta_{\bSigma}^* &= \text{argmax}_{\btheta_{\bSigma}} \left\{ - \frac{T}{2} \log |\bSigma(\btheta_{\bSigma})| - \frac{1}{2} \text{tr} \left\{ \bSigma(\btheta_{\bSigma})^{-1} \sum_{t=1}^T \langle(\bseta_{0,t} \odot \sqrt{\bZ_{t}}) (\bseta_{0,t} \odot \sqrt{\bZ_{t}} ) ^\top \rangle \right\} \right\},
 \end{align*}
 otherwise,
\[ \hat{\bSigma}^{(k+1)} = \frac{1}{T} \sum_{t=1}^T \langle(\bseta_{0,t} \odot \sqrt{\bZ_{t}}) (\bseta_{0,t} \odot \sqrt{\bZ_{t}} ) ^\top \rangle.  \]


\section{Simulation Study}\label{sec:SimStudy}

We conduct a simulation study in order to assess the relative performance of the proposed model, which we denote at SKT, and a subclass where there is no skew in the fine-scale dependence, which we refer as the \textsl{Gaussian} model, or GAU. 

GAU assumes that $Z_r=1$ in distribution, and $\bsy{\delta}_r=\mbf{0}$ for all $r$ in \eqref{eq:Yr_rep}, so that the expression becomes $\bY_r  = \bzeta_r \bX_0 + \bDelta_{\zeta_r}\bU_r$, where\footnote{with an abuse of notation to considerably simplify the exposition, the parameters for GAU will have the same notation as the parameters for SKT} $\mbf{U}_r \sim N_{d_r}(\mbf{0},\bsy{\Psi}_r)$, and $\bX_0 \sim N_R(\mbf{0},\bsy{\Sigma})$. For computational convenience, we consider $\bzeta_r = \zeta_r \1_{d_r}$. Thus, the model assumes a combined effect of a fine and large-scale effect, whose relative contribution is modulated by $\zeta_r \in (-1,1)$.  Therefore, for GAU, $\bsy{\theta}=\left(\{\zeta_r,\bsy{\theta}_{\bsy{U}_r}^{\top}, r=1, \ldots, R\},\bsy{\theta}_{\bX_0}^{\top}\right)^{\top}$, where as before $\bsy{\Psi}_r=\bsy{\Psi}_r(\bsy{\theta}_{\bsy{U}_r})$ and $\bsy{\Sigma}=\bsy{\Sigma}(\bsy{\theta}_{\bX_0})$, while the only latent process is $\bX_0$.

The joint distribution at time $t$ is a simplified version of \eqref{eq:joint}, and is given by
\begin{equation*}
\begin{array}{lll}
p( \by_t, \bX_{0,t} | \btheta) & = & p(\by_t | \bX_{0,t},  \btheta) p(\bX_{0,t} |  \btheta) \\[7pt]
& = & \prod_{r=1}^R \frac{1}{(2\pi)^{d_r/2}} \frac{1}{(1-\zeta_r^2)^{d_r/2} | \bPsi_r|^{1/2}} \exp\left\{
-\frac{1}{2(1-\zeta_r^2)} (\by_r - \zeta_r\1_{d_r} X_{0,r,t})^\top \bPsi_r^{-1} (\by_r - \zeta_r\1_{d_r} X_{0,r,t})\right\} \\[7pt]
&   & \times\frac{1}{(2\pi)^{R/2}} \frac{1}{|\bSigma|^{1/2}} \exp\left\{-\frac{1}{2}\bX_{0,t}^\top \bSigma^{-1}\bX_{0,t}  \right\}.
\end{array}
\end{equation*}

If we denote by $\mbf{y}=(\mbf{y}_1^{\top}, \ldots, \mbf{y}_T^{\top})^{\top}$ and $\mbf{X}_0=(\mbf{X}_{0,1}^{\top}, \ldots, \mbf{X}_{0,T}^{\top})^{\top}$, the log-likelihood for all time points, excluding additive constants, is
\begin{equation*}
\boldsymbol{\ell}(\btheta | \by, \bX_0)  = \sum_{r=1}^R \boldsymbol{\ell}_r(\btheta | \by, \bX_0)   - \frac{T}{2}\log|\bSigma|    -  \frac{1}{2} \sum_{t=1}^T \bX_{0,t}^\top \bSigma^{-1} \bX_{0,t},
\end{equation*}
with 
\begin{equation*}
\begin{split}
\boldsymbol{\ell}_r(\btheta | \by, \bX_0)  =  & -\frac{Td_r}{2}\log (1-\zeta_r^2) - \frac{T}{2} \log |\bPsi_r| \\ 
& - \frac{1}{2(1-\zeta_r^2)}\sum_{t=1}^T (\by_r - \zeta_r\1_{d_r} X_{0,r,t})^\top \bPsi_r^{-1} (\by_r - \zeta_r\1_{d_r} X_{0,r,t}). 
\end{split}
\end{equation*}

An EM algorithm is again used for inference. The E-step requires the computation of $\mathbb{E}(X_{0,r,t}| \by, \btheta)$ and $\mathbb{E}(X_{0,r,t}^2 | \by, \btheta)$, which can be easily obtained by Gibbs sampling from the joint distribution $p(\mbf{X}_{0,t} | \by, \btheta)$ as the full conditionals are available in closed form. Indeed, for  the case $R=2$, $p(X_{0,1,t}| X_{0,2,t}, \by, \btheta) \sim  N(\mu_{0,1},\sigma_{0,1}^2)$, with
\[
\mu_{0,1} = \frac{ \frac{\hat{\zeta}_r}{1-\hat{\zeta}_r^2} \1_{d_r}^\top \hat{\bPsi}_r^{-1} \by_r - X_{0,2} \hat{\lambda}_{1,2}}{\frac{\hat{\zeta}^2_r}{1-\hat{\zeta}_r^2}\1_{d_r}^\top \hat{\bPsi}_r^{-1}\1_{d_r} + \hat{\lambda}_{r,r}^2}, \qquad \sigma^2_{0,1} = \frac{1}{\frac{\hat{\zeta}^2_r}{1-\hat{\zeta}_r^2}\1_{d_r}^\top \hat{\bPsi}_r^{-1}\1_{d_r} + \hat{\lambda}_{r,r}^2}. 
\]

\noindent The M-step consists of the following updates:\\

\noindent 1) \textbf{Update $\bPsi_r^{(k)}$:} 
The non-parametric update is as follows,
\begin{equation*}
\hat{\bPsi}_r^{(k+1)} = \frac{1}{T(1-\hat{\zeta_r}^2)} \hat{\bB}_r, 
\end{equation*}
with
\[ \hat{\bB}_r =  \sum_{t=1}^T \by_{r,t} \by_{r,t}^\top - \hat{\zeta}_r \sum_{t=1}^T \langle X_{0,r,t} \rangle \by_{r,t} \1_{d_r}^\top -  \hat{\zeta}_r \sum_{t=1}^T \langle X_{0,r,t} \rangle \1_{d_r} \by_{r,t}^\top + \hat{\zeta}_r^2 \sum_{t=1}^T \langle X_{0,r,t}^2 \rangle  \1_{d_r}  \1_{d_r}^\top. \]
In the parametric case, with $\bPsi_r= \bPsi_r(\btheta_{\bPsi_r})$, we have
\[ \hat{\bPsi}_r^{(k+1)}  =  \hat{\bPsi}_r^{(k+1)} (\btheta_{\bPsi_r}^*), \quad \btheta_{\bPsi_r}^*= \text{argmax}_{\btheta_{\bPsi_r}} \left\{ - \frac{T}{2} \log |\bPsi_r(\btheta_{\bPsi_r})| - \frac{1}{2(1-\zeta_r^2)} \text{tr} \left\{ \bPsi_r(\btheta_{\bPsi_r})^{-1}  \hat{\bB}_r \right\}\right\}.\] 


\noindent 2) \textbf{Update $\zeta_r^{(k)}$:} The update is obtained numerically, 
\begin{equation*}
 \hat{\zeta}_r^{(k+1)} = \text{argmax}_{\zeta_r} \left\{ - \frac{Td_r}{2}\log (1-\zeta_r^2) - \frac{1}{2(1-\zeta_r^2)} \text{tr} \left\{ \hat{\bPsi}_r^{-1}  \hat{\bB}_r \right\} \right\}.
\end{equation*}
 \vspace{.1in}
 
\noindent 3) \textbf{Update $\bSigma^{(k)}$:} As above, in the non-parametric setting, the update would be given by
\[ \hat{\bSigma}^{(k+1)} = \frac{1}{T} \langle \bX_{0,t} \bX_{0,t}^\top  \rangle,\]
whereas in the parametric setting,
\[ \hat{\bSigma}^{(k+1)}  =  \hat{\bSigma}^{(k+1)} (\btheta_{\bSigma}^*), \quad \btheta_{\bSigma}^*= \text{argmax}_{\btheta_{\bSigma}} \left\{ - \frac{T}{2}\log|\bSigma(\btheta_{\bSigma})|    -  \frac{1}{2} \text{tr} \left\{ \bSigma(\btheta_{\bSigma})^{-1} \sum_{t=1}^T  \langle \bX_{0,t} \bX_{0,t}^\top \rangle \right\} \right\}.\]


\noindent We generate 100 simulations with $T=1,000$ from SKT, and fit both SKT and GAU according to their respective EM algorithms, considering $M=500$ iterations for each time-step $t$. The simulation considers 2 regions, with 9 points each arranged in a $3\times 3$ regular grid. The scale matrices for both models are parametrized according to a Mat{\'e}rn correlation function for distance $h$ defined as $(h/\phi)^{\nu} \mathcal{K}_{\nu}(h/\phi)$, where $\mathcal{K}_{\nu}(\cdot)$ is a Bessel function of the second kind, $\phi$ is the range, while $\nu$ controls the differentiability of the sample paths of the spatial stochastic process \citep{ste99}. Here we fix $\nu=1.5$, implying that the sample paths are one time differentiable, a plausible assumption given the smoothness of the wind field data. See the Appendix for further details regarding the chosen parameter values. 

\begin{figure}[htbp]\centering
\includegraphics[width=3.1in]{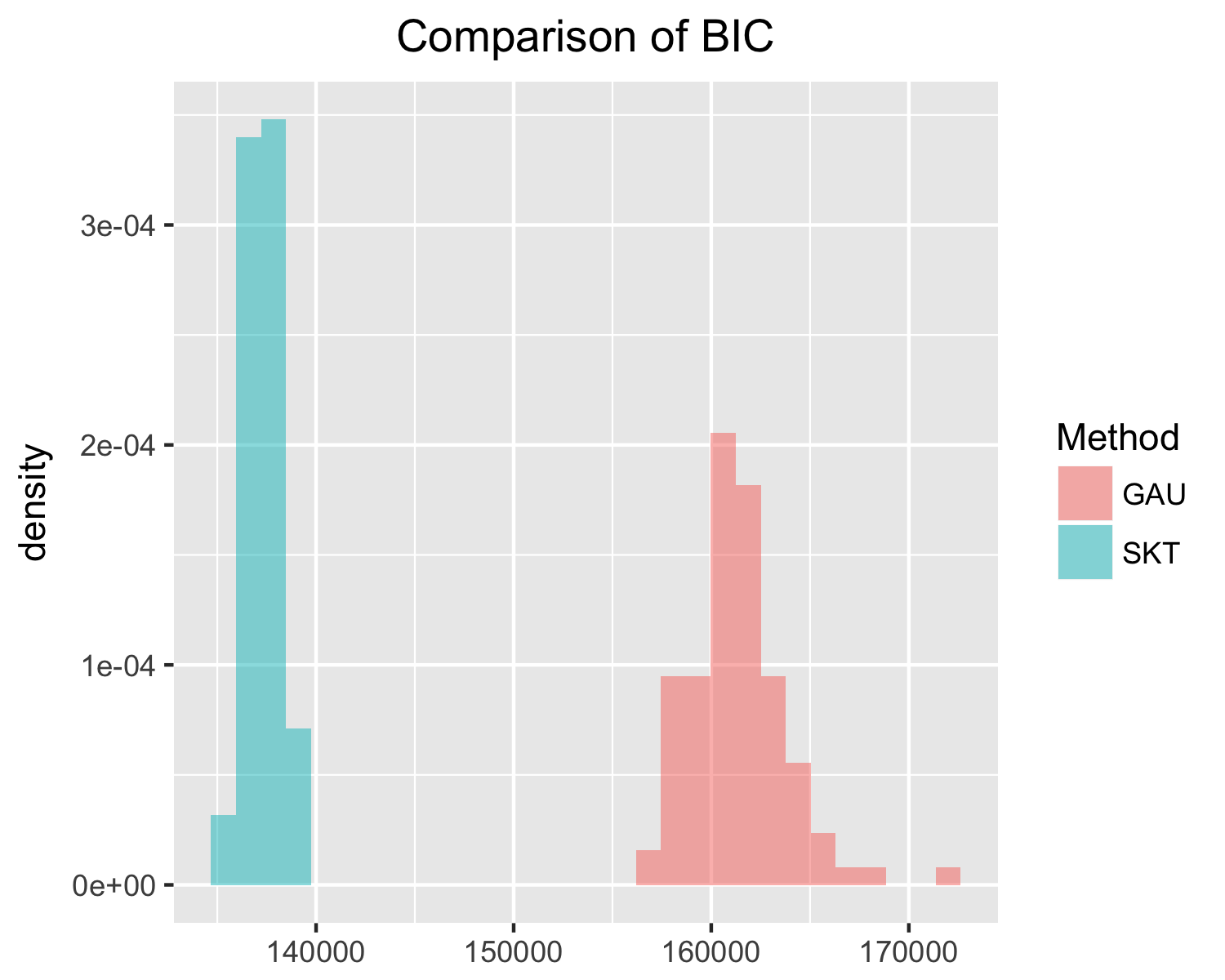}
\includegraphics[width=3.1in]{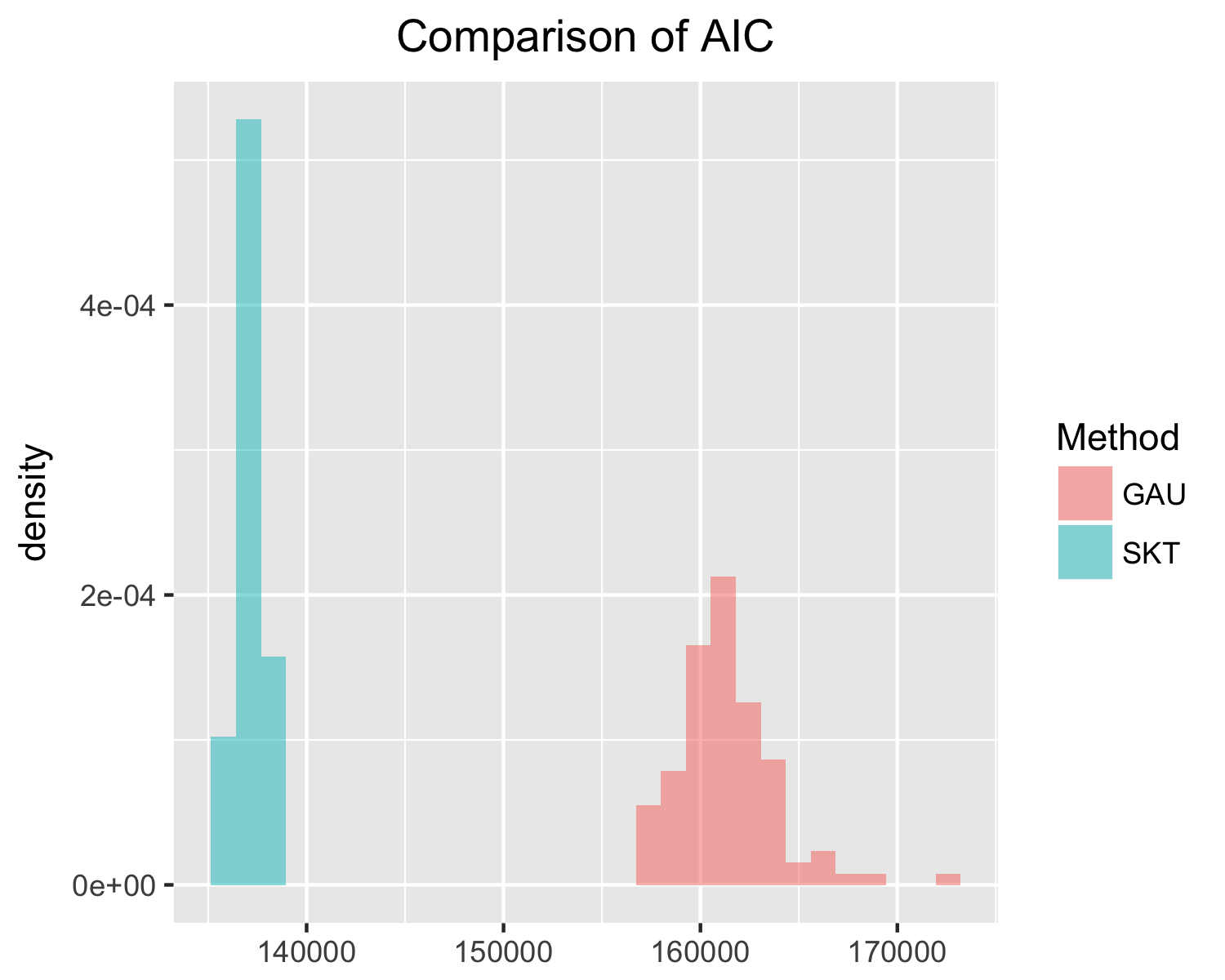}
\caption{Comparison of SKT and GAU in terms of BIC and AIC values from 100 simulations, assuming SKT is the true model.}\textbf{•}
\label{fig:BIC}
\end{figure}

Figure \ref{fig:BIC} presents the Bayesian Information Criterion (BIC) and Akaike Information Criterion (AIC), for GAU and SKT. SKT clearly outperforms GAU by a considerable extent according to both indices, and uniformly across simulations. Hence, the simpler Gaussian model GAU with restriction $Z_r=1$ in distribution, and $\bsy{\delta}_r=\mbf{0}$ for all $r$ in \eqref{eq:Yr_rep} results in a considerable misfit.


\section{Application to Wind Data}\label{sec:Application}

Stochastic generators \citep{cas17,jeo17a,jeo17b} are statistical models designed to approximate climate model output run under different initial conditions but same scenario and physical parameters. Once the SG has been trained with a small number of runs, it can generate instantaneous surrogate runs that are able to reproduce the extent of the variability of a variable without resorting to additional expensive simulations.  

Of particular importance here is the study of the variability of wind speed in Saudi Arabia at policy-relevant scale (i.e., daily level). \citet{tagle2017assessing} introduced a new class of models, which we hereby denote as SKT-I, for daily wind speed, where the residuals of a vector-autoregressive mean structure were modeled regionally using a multivariate skew-$t$ distribution. Although the model adequately captured several features of the process, it made the unrealistic assumption that each region of the spatial domain evolved independently. 

By applying the model described in Section \ref{sec:model}, we extend the work by introducing dependence across regions by means of a large-scale effect. The wind data is provided by the Large Ensemble project developed at the National Center for Atmospheric Research (NCAR), which consists of a collection of 30 simulations of past and future climate on a global scale at approximately $1^\circ$ resolution \citep{kay15}.


Following the notation of \citet{tagle2017assessing}, we assume that $\mathbf{W}_{t} = (W_{1,t},\ldots,W_{d,t})^\top$ is the $d$-vector of daily wind speeds over the domain at day $t$, SKT-I assumes that $\bW_{t} = \bmu_{t} + \bsy{\Xi}_t^{1/2} \bvepsilon_{t}$, for $t=1,\ldots,T$,
where $\bmu_{t}$ represents a vector-autoregressive (VAR) process at time $t$, $\bsy{\Xi}_t^{1/2}$ is a $d \times d$ time-varying diagonal matrix of scaling factors at time $t$, and $\bvepsilon_{t}$ an innovation process. The authors found that a VAR(2) specification with a first-order stencil neighborhood scheme for the first matrix and a diagonal form for the second adequately represents the temporal and cross-temporal structure. Here $\bsy{\Xi}_t^{1/2}$ is estimated by regressing the standard deviation of the residuals to a set of harmonics for each calendar day over all years.  In order to account for spatial non-stationarity, the vector $\bvepsilon_{t}$ is partitioned into subvectors $\bvepsilon_{r,t}$, $r=1,\ldots,R$, with $r$ denoting a particular region of the spatial domain. The cardinality of the regions varies between 8 and 27 points, and each $\bvepsilon_{r,t}$ follows an independent multivariate skew-$t$ distribution designed to capture the small-scale features of the wind speed field.

We now apply the model outlined in Section \ref{sec:model} (consistently with Section \ref{sec:SimStudy}, we denote it as SKT), and we use a Mat{\'e}rn correlation function for both $\bSigma$ and $\bsy{\Psi}_r$, $r=1,\ldots,R$, with a fixed value of $\nu = 1.5$. For the former, the regional centroid distances are considered. Based on exploratory analysis, the range parameter for $\bsy{\Psi}_r$ is fixed at $\phi=100$, while for $\bSigma$ it is set at 1000; $\nu_r = 3$, and $\bzeta_r = 0.3 \cdot \1_{d_r}$ across all regions; lastly, the values of $\bdelta_r$ are obtained from the cited paper, subject to the transformation from the direct parameterization (DP) to the centered parameterization (CP) \citep{azzalini1996multivariate}. 

We quantify the added benefit of a large scale effect by computing the relative improvement in the estimation of the covariance using SKT or SKT-I. If we denote by $\hat{C}_{\text{SKT}}$ ($\hat{C}_{\text{SKT-I}}$) the covariance among regions implied by model SKT (SKT-I), and by $\| \cdot \|_F$ the Frobenius norm, the relative improvement is $\|\hat{C}_{\text{SKT}}-\hat{C}_{\text{SKT-I}}\|_F/\|\hat{C}_{\text{SKT-I}}\|_F\approx 0.23$.

\begin{figure}[htbp]\centering
\includegraphics[width=4.5in]{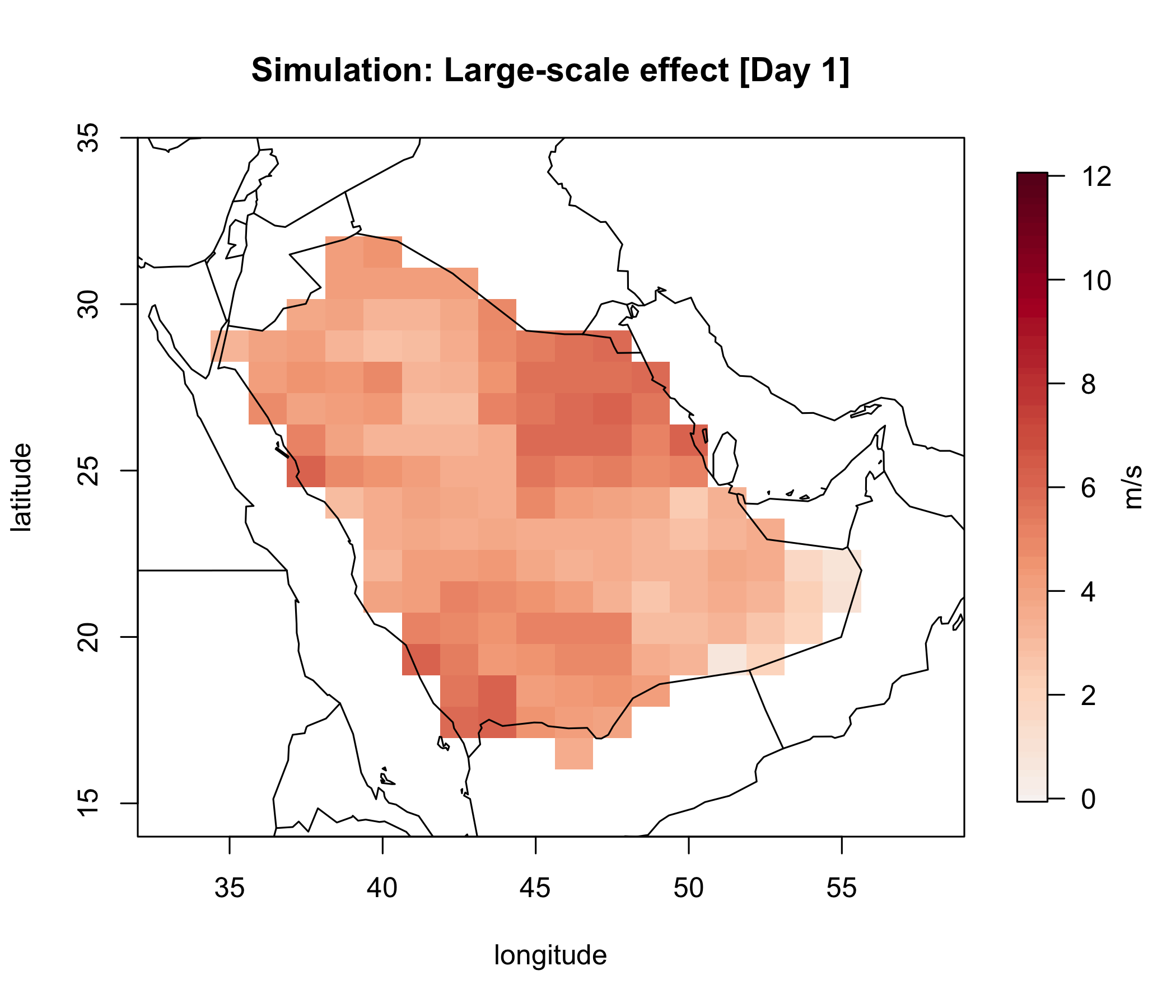}
\caption{A visuanimation \citep{gen15} of the wind fields from the SG (in m/s) in January 2004 from the proposed model. (available upon demand)}\label{M2}
\end{figure}

Movie \ref{M2} reports a visuanimation \citep{gen15} of a surrogate run from the SG for the month of January 2004. The movie shows spatially coherent fields with no apparent discontinuity across regions, consistently with the runs from the original data set (see supplementary material).

\renewcommand{\figurename}{Movie}

\section{Conclusions}\label{sec:Conclusions}

This study introduced a multi-resolution modeling framework based on the ST family of distributions, which are characterized by their flexibility in representing asymmetry and heavy-tails. The additional flexibility comes at the expense of analytical tractability, in particular, closure under convolutions, thus the model construction relies on the SN from which the ST distribution is derived. Accordingly, what may be interpreted as a large-scale effect is assumed to follow a multivariate normal distribution, while regional fine-scale effects follow a multivariate SN, allowing for the overall regional effects to be distributed according to a multivariate ST distribution. This leads to a convenient modeling scheme where the regions are allowed to evolve independently, while the dependence across regions is captured in the covariance function of the large-scale effects. Inference is based on the maximization of the likelihood, for which a Monte Carlo EM algorithm was developed. 

Extending the work of \citet{tagle2017assessing}, the proposed model was fit to the time series of daily wind speed residuals over Saudi Arabia. While the presence of the large-scale effect enhanced the degree of spatial dependence, its magnitude is lower than that implied by the empirical estimates. The lack of flexibility in capturing higher degrees of spatial dependence can be explained by a trade-off in the model construction between the representation of the higher-order moments of the marginal distribution and the degree of spatial correlation. More specifically, for any two points, $p =1,2,$ belonging to two different regions, $r_1$ and $r_2$, we have that the  correlation between their responses $ \text{Cor}(Y_1,Y_2) \propto \zeta_{r_1} \zeta_{r_2} \sigma_{r_1,r_2}$, where $\sigma_{r_1,r_2}$ corresponds to the correlation between both regions as represented by the correlation matrix $\bSigma$. Thus, the smaller the magnitude of $\zeta_{r_1}$, the smaller the correlation $\text{Cor}(Y_1,Y_2)$, but at the same time, the smaller the weight of $X_{0,r_1}$ in the presentation of $Y_1$, and consequently, the greater the degree of skewness of the resulting marginal distribution. Future work will be aimed at constructions that allow for greater flexibility in capturing spatial correlations. Nonetheless, the proposed framework allows for a meaningful increase in the size of the spatial domains that can be tractably analyzed, thereby overcoming an important hurdle in the widespread adoption of this family of distributions in the study of geophysical processes. 
 
\section*{Appendix}
{\large \textbf{Proof of Proposition 1.} }\\
For ease of notation, we prove the skew-normal case where $Z=1$, the general proof follows from the definition of the skew-$t$ distribution. While the general result of the sum of a multivariate normal random variate and a SN random variate is well-known to belong to the latter family of distributions, the purpose of this proof is to provide explicit expressions for the parameters of the SN distribution that arises from the proposed linear combination. The derivation follows closely that of the additive representation of the SN distribution found in the appendix of \citet{azzalini1996multivariate}. For notational convenience, let $u_j = y_j/(1-\zeta_j^2)^{1/2}$, $j=1,\ldots,d$, and $\bu= (u_1,\ldots,u_d)^{\top}$. Using standard  transformation methods of random variables, the density function of $\bY$ at point $\by \in \mathbb{R}^d$, is
\begin{align*}
f_z(z) =&  \int_\mathbb{R} 2 \phi_d(\bu - \blambda v; \bOmega) \Phi(\balpha^\top (\bu - \blambda v)) \phi(v) \frac{1}{\prod_{i=1}^d (1-\zeta_i^2)^{1/2}}\mathrm{d}v\\
		  =&   \int_\mathbb{R}  \frac{2}{(2\pi)^{d/2}|\bOmega|^{1/2} \prod_{i=1}^d (1-\zeta_i^2)^{1/2}} e^{-\frac{1}{2} (\bu - \blambda v)^\top \bOmega^{-1}   (\bu - \blambda v)} \frac{1}{\sqrt{2\pi}} e^{\frac{-v^2}{2}} \Phi(\balpha^\top (\bu - \blambda v))\mathrm{d}v \\
		  =&   \int_\mathbb{R}  \frac{2}{(2\pi)^{d/2}|\bOmega|^{1/2}\prod_{i=1}^d (1-\zeta_i^2)^{1/2}} e^{-\frac{1}{2}\left( \bu^\top \bOmega^{-1} \bu - \frac{(\blambda^\top \bOmega^{-1} \bu)^2}{1+\blambda^\top \bOmega^{-1} \blambda} \right)}   \\
		 & \times  \frac{1}{\sqrt{2\pi}} e^{-\frac{1}{2}  \left[ (1+\blambda^\top \bOmega^{-1} \blambda)^{1/2} \left(v - \frac{\blambda^\top \bOmega^{-1} \bu}{1+\blambda^\top \bOmega^{-1} \blambda} \right)\right]^2}\Phi(\balpha^\top (\bu - \blambda v))\mathrm{d}v \\
		  = &  \frac{2}{ (2\pi)^{d/2}|\bOmega|^{1/2}(1+\blambda^\top \bOmega^{-1} \blambda)^{1/2}\prod_{i=1}^d (1-\zeta_i^2)^{1/2}} e^{-\frac{1}{2}\left( \bu^\top \bOmega^{-1} \bu - \frac{(\blambda^\top \bOmega^{-1} \bu)^2}{1+\blambda^\top \bOmega^{-1} \blambda} \right)}  \\
		 & \times  \int_\mathbb{R} \phi(t) \Phi \left[ \balpha^\top  \left( \bu - \blambda \left( \frac{t}{ (1+\blambda^\top \bOmega^{-1} \blambda)^{1/2}} +   \frac{\blambda^\top \bOmega^{-1} \bu}{1+\blambda^\top \bOmega^{-1} \blambda} \right) \right) \right] \mathrm{d}t.
\end{align*}
Using the binomial inverse theorem, as in \citet{mardia1980multivariate} (A.2.4f), 
\begin{align*}
 \bu^\top \bOmega^{-1} \bu - \frac{(\blambda^\top \bOmega^{-1} \bu)^2}{1+\blambda^\top \bOmega^{-1} \blambda} & = \bu^\top \left( \bOmega^{-1}  -  \frac{ \bOmega^{-1} \blambda \blambda^\top \bOmega^{-1}}{1+\blambda^\top \bOmega^{-1} \blambda} \right ) \bu \\
 & =  \bu^\top \left( \bOmega   + \blambda \blambda^\top \right )^{-1} \bu \\
 & =  \bz^\top \bDelta^{-1} \left( \bOmega   + \blambda \blambda^\top \right )^{-1} \bDelta^{-1}  \bz \\
 & =  \bz^\top \left( \bDelta (\bOmega   + \blambda \blambda^\top) \bDelta \right )^{-1}  \bz = \bz^\top \bOmega^*  \bz.
\end{align*} 
Furthermore
\[ | \bOmega^*| =  |\bOmega|(1+\blambda^\top \bOmega^{-1} \blambda)\left(\prod_{i=1}^d (1-\zeta_i^2)^{1/2}\right)^2. \]
Thus the first term in the above equation reduces to $2 \phi_d(\by; \bOmega^*)$. Lastly, note that the argument of the distribution function $\Phi$ can be written as
\begin{align*}
&\balpha^\top  \left( \bu - \blambda \left( \frac{t}{ (1+\blambda^\top \bOmega^{-1} \blambda)^{1/2}} +   \frac{\blambda^\top \bOmega^{-1} \bu}{1+\blambda^\top \bOmega^{-1} \blambda} \right) \right) \\
& =  -\frac{\balpha^\top \blambda }{(1+\blambda^\top \bOmega^{-1} \blambda)^{1/2}}t + \balpha^\top   \left( \bu - \frac{\blambda \blambda^\top \bOmega^{-1} \bu}{1+\blambda^\top \bOmega^{-1} \blambda} \right) \\
 &=  -\frac{\balpha^\top \blambda }{(1+\blambda^\top \bOmega^{-1} \blambda)^{1/2}}t + \balpha^\top \bOmega  \left( \bOmega^{-1} - \frac{\bOmega^{-1}\blambda \blambda^\top \bOmega^{-1} }{1+\blambda^\top \bOmega^{-1} \blambda} \right)\bu 
\end{align*}
\begin{align*}
& =  -\frac{\balpha^\top \blambda }{(1+\blambda^\top \bOmega^{-1} \blambda)^{1/2}}t + \balpha^\top \bOmega  \left( \bOmega   + \blambda \blambda^\top \right )^{-1} \bu  \\
& =  -\frac{\balpha^\top \blambda }{(1+\blambda^\top \bOmega^{-1} \blambda)^{1/2}}t + \balpha^\top \bOmega  \bDelta {\bOmega^*}^{-1} \by.
\end{align*}
Therefore, the integral is of the form $\mathbb{E}[\Phi(hT-k)]$, with $T \sim N(0,1)$, and applying Lemma 2.2 in \citet{azzalini2014skew} yields 
\begin{align*}
\mathbb{E} \left[  -\frac{\balpha^\top \blambda }{(1+\blambda^\top \bOmega^{-1} \blambda)^{1/2}}T+ \balpha^\top \bOmega  \bDelta {\bOmega^*}^{-1} \by\right] & = \Phi \left[\left( 1 + \frac{\balpha^\top \blambda \blambda^\top \balpha}{1 + \blambda^\top \bOmega^{-1} \blambda} \right)^{-1/2} \balpha^\top \bOmega  \bDelta {\bOmega^*}^{-1} \by \right].
\end{align*}

\vspace{.5in}
\noindent {\large \textbf{Simulation Study: parameter values}}\\
The distances of the $3 \times 3$ regular grid were based on the approximate $1 \times 1$ longitude-latitude grid used in the NCAR climate model output. The latitudes ranged between $20.26^\circ$ and $22.15^\circ$. For the Mat{\'e}rn parameterization of the scale matrices, the range parameter $\phi$ was fixed at 90, and the distances were computed using the \verb|rdist.earth| from the \verb|fields| R package. The values of $\delta_r$, $r=1,2$, were chosen to be equal, and based on the estimates obtained in \citet{tagle2017assessing}, with values ranging between $0.453$ and $0.754$; $\zeta_1 = \zeta_2 = 0.2$, $\nu_1 = \nu_2 = 3$, and the off-diagonal element of $\bSigma$ was fixed to $0.9$.



\baselineskip 23 pt
\bibliographystyle{agsm}
\bibliography{notes_references}

\end{document}